\documentclass[english,a4paper,pra,twocolumn,superscriptaddress,showpacs,preprintnumbers]{revtex4}
\usepackage{graphicx}
\usepackage{amssymb}

\makeatletter

\makeatletter

\usepackage{dcolumn}

\usepackage{bm}

\newcommand{\be}{\begin{equation}}
\newcommand{\ee}{\end{equation}}
\newcommand{\bae}{\begin{eqnarray}}
\newcommand{\eae}{\end{eqnarray}}

\newcommand{\ket}[1]{| #1 \rangle}

\makeatother

\makeatother

\usepackage{babel}
\makeatother
\begin{document}

\title{Bures metric over thermal state manifolds and quantum criticality}

\author{Paolo Zanardi}

\affiliation{Department of Physics and Astronomy, University of Southern California
Los Angeles, CA 90089-0484 (USA) }

\affiliation{Institute for Scientific Interchange, Villa Gualino, Viale Settimio
Severo 65, I-10133 Torino, Italy }

\author{Lorenzo Campos Venuti}

\affiliation{Institute for Scientific Interchange, Villa Gualino, Viale Settimio
Severo 65, I-10133 Torino, Italy }

\author{Paolo Giorda}

\affiliation{Institute for Scientific Interchange, Villa Gualino, Viale Settimio
Severo 65, I-10133 Torino, Italy }

\date{\today}

\begin{abstract}
We analyze the Bures metric over the manifold of thermal density matrices
for systems featuring a zero temperature quantum phase transition.
We show that the quantum critical region can be characterized in terms
of the temperature scaling behavior of the metric tensor itself. Furthermore,
the analysis of the metric tensor when both temperature and an external
field are varied, allows to complement the understanding of the phase
diagram including cross-over regions which are not characterized by
any singular behavior. These results provide a further extension of
the scope of the metric approach to quantum criticality.
\end{abstract}


\maketitle

\section{Introduction}

These years are witnessing an increasing research effort at the intersection
of Quantum Information Science \cite{qis} and more established fields
like theoretical condensed matter \cite{qpt-qis}. It belongs to this
class the approach to quantum phase transitions (QPT) \cite{sachdev}
based on the information-geometry of quantum states that has been
recently proposed in Ref \cite{za-pa} and \cite{zhou}. Further developments,
for specific, yet important, class of quantum states have been then
reported in \cite{za-co-gio,co-gio-za,co-ion-za,pier,hamma,gu,zhou1,Gu}.
The underlying idea is deceptively simple: the major structural change
in the ground state (GS) properties at the QPT should reveal itself
by some sort of singular behavior in the distance function between
the GSs corresponding to slightly different values of the coupling
constants. This intuition can me made more quantitative by analyzing
the leading order terms in the expansion of the {\em quantum fidelity}
between close GSs.

A general differential-geometric framework encompassing all of these
result has been offered in Ref \cite{DG-qpt}. There it has been shown
that these leading order terms do correspond to a Riemannian metric
$g$ over the parameter manifold. This metric $g$ is nothing but
the pull-back of the natural metric over the projective Hilbert space
via the map associating the Hamiltonian parameters with the corresponding
GS. In the thermodynamical limit the singularities of $g$ correspond
to QPTs. In Ref. \cite{lor} the nature of this correspondence has
been further investigated and it has been shown that both the metric
approach to QPT and the one based on geometrical phases \cite{BP-qpt0,BP-qpt1}
can be understood in terms of the critical scaling behavior of the
quantum geometric tensor \cite{pro}.

The conceptually appealing and potentially practically relevant feature
of this strategy consists of the fact that its viability does not
rely on any a priori knowledge of the physics of the model e.g., order
parameters, symmetry breaking patterns,... but just on a universal
geometrical structure (basically the Hilbert scalar product). Very
much in the spirit of Quantum Information the metric approach is fully
based on quantum states rather than Hamiltonians (that might be even
unknown), once these are given the machinery can be applied.

In this paper we further extend the scope of this metric approach
by considering the manifold of thermal states of a family of Hamiltonians
featuring a zero-temperature PT. In \cite{zhong-guo} it was shown
that by studying the mixed-state fidelity \cite{Uhlmann} between
Gibbs states associated with slightly different Hamiltonians one could
detect the influence of the zero-temperature quantum criticality over
a finite range of temperatures. Here we will refine that analysis
and make it more quantitative by resorting to the concept of {\em
Bures} metric between mixed quantum states. This metric provides
the natural finite-temperature extension of the metric tensor $g$
studied in the GS case and corresponds again to the leading order
in the expansion of the (mixed-state) fidelity between close states
i.e., associated with infinitesimally close parameters. By analyzing
the case of the Quantum Ising model we shall show how the quantum-critical
region above the zero-temperature QPT can be remarkably characterized
in terms of the scaling behavior of the Bures metric tensor.

The paper is organized as follows: in sect. II we introduce the basic
concepts about mixed-state metrics and in Sect III we specialize them
to the case of thermal (Gibbs) states. In sect. IV we provide generalities
about quasi-free fermion systems and in Sect VI we analyze in detail
the Bures metric tensor for the quantum Ising model. Finally in sect
V conclusions and outlook are given.

\section{Preliminaries}

The Bures distance between two mixed-states $\rho$ and $\sigma$
is given in terms of the Uhlmann fidelity \cite{Uhlmann} \begin{equation}
{\cal F}(\rho,\sigma)={\rm {tr}\sqrt{\rho^{1/2}\sigma\rho^{1/2}}}\label{fid}\end{equation}
 by %$d_{B}(\rho,\sigma)=\cos^{-1}{\cal F}(\rho,\sigma)$
$d_{B}(\rho,\sigma)=\sqrt{2\left[1-{\cal F}(\rho,\sigma)\right]}$.

The starting point of our analysis is provided by the following expression
for the Bures distance between two infinitesimally close density matrices
(see e.g.~\cite{bures-formula} for a derivation) \begin{equation}
ds^{2}(d\rho):=d_{B}^{2}\left(\rho,\rho+d\rho\right)=\frac{1}{2}\sum_{n,m}\frac{\left|\langle m|d\rho|n\rangle\right|^{2}}{p_{m}+p_{n}},\label{eq:dist}\end{equation}
 where $|n\rangle$ is the eigenbasis of $\rho$ with eigenvalues
$p_{n}$ i.e., $\rho=\sum_{n}p_{n}|n\rangle\langle n|.$ Even though
in the sum in (\ref{eq:dist}) $p_{n}$ and $p_{m}$ cannot be simultaneously
in the kernel of $\rho,$ since $|n\rangle,|m\rangle\in{\rm {Ker}}(\rho)\Rightarrow\langle n|d\rho|m\rangle=0,$
one can formally extend the sum to all possible pairs by setting to
zero the unwanted terms. For $\rho$ pure i.e., $\rho=|\psi\rangle\langle\psi|$
one has $d\rho=|d\psi\rangle\langle\psi|+|\psi\rangle\langle d\psi|$
from which one sees that the diagonal matrix elements of $d\rho$
are vanishing and one is left with $ds_{B}^{2}=\sum_{m\in{\rm {Ker}(\rho)}}|\langle d\psi|m\rangle|^{2}=\langle d\psi|(1-|\psi\rangle\langle\psi|)|d\psi\rangle.$
This expression coincides with the Riemannian metric considered in
\cite{DG-qpt}. The bures metric (\ref{eq:dist}) is tightly connected
to the so-called quantum Fisher information and it appears in the
quantum version of the celebrated Cramer-Rao bound \cite{bra-ca}.
This suggests the possible relevance of the results that we are going
to present in this paper to the field of quantum estimation \cite{estima}.

To begin with we would like to cast Eq. (\ref{eq:dist}) in a from
suitable for future elaborations. Let us first differentiate the density
matrix $d\rho=\sum_{n}(dp_{n}|n\rangle\langle n|+p_{n}|dn\rangle\langle n|+p_{n}|n\rangle\langle dn|)$
and consider to begin the matrix element $(d\rho)_{ij}$. We observe
that $\langle i|j\rangle=\delta_{i,j}\Rightarrow\langle di|j\rangle=-\langle i|dj\rangle;$
whence $\langle i|d\rho|j\rangle=\delta_{i,j}dp_{i}+\langle i|dj\rangle(p_{j}-p_{i}).$
Putting this expression back into (\ref{eq:dist}) one obtains \begin{equation}
ds^{2}=\frac{1}{4}\sum_{n}\frac{dp_{n}^{2}}{p_{n}}+\frac{1}{2}\sum_{n\neq m}|\langle n|dm\rangle|^{2}\frac{(p_{n}-p_{m})^{2}}{p_{n}+p_{m}}\ \ .\label{start}\end{equation}

This relation is quite interesting since it tells apart the classical
and the quantum contributions. Indeed the first term in (\ref{start})
is nothing but the {\em Fisher-Rao} distance between the probability
distributions $\{ p_{n}\}_{n}$ and $\{ p_{n}+dp_{n}\}_{n}$ whereas
the second term takes into account the generic non-commutativity of
$\rho$ and $\rho^{\prime}:=\rho+d\rho.$ We will refer to these two
terms as the classical and non-classical one respectively. When $[\rho^{\prime},\,\rho]=0$
the problem gets effectively classical and the Bures metric collapses
to the Fisher-Rao one; this latter being in general just a lower bound
\cite{bra-ca,Woo}.

Before moving to the analysis of the metric (\ref{eq:dist}) we would
like to comment about the connection with the recently established
quantum Chernoff bound \cite{QCB}. This latter, denoted by $\xi_{QCB},$
is the quantum analogue of the Chernoff bound in classical information
theory; it quantifies the rate of exponential decay of the probability
of error in discriminating two quantum states $\rho$ and $\sigma$
when a large number $n$ of them is provided and collective measurements
are allowed i.e., $P_{err}\sim\exp(-n\xi_{QCB}).$ The Chernoff bound
naturally induces a distance function over the manifold of quantum
states with a well defined operational meaning (the bigger the distance
between the states the smaller the asymptotic error probability in
telling one from the other). In \cite{QCB} it has been proven that
$\exp(-\xi_{QCB})=\min_{0\le s\le1}{\rm {tr}\left(\rho^{s}\sigma^{1-s}\right)\le{\cal F}(\rho,\sigma)}$
and that for infinitesimally close states i.e., $\sigma=\rho+d\rho,$
one has \begin{equation}
ds_{QCB}^{2}:=1-\exp(-\xi_{QCB})=\frac{1}{2}\sum_{n,m}\frac{\left|\langle m|d\rho|n\rangle\right|^{2}}{(\sqrt{p_{m}}+\sqrt{p_{n}})^{2}}.\label{QCB-metric}\end{equation}
 From this expression we see that the distinguishability metric associated
with the quantum Chernoff bound has the same form of the Bures one
(\ref{eq:dist}) but the denominators $p_{n}+p_{m}$ are replaced
by $(\sqrt{p_{m}}+\sqrt{p_{n}})^{2}.$ Using the inequalities $(\sqrt{p_{m}}+\sqrt{p_{n}})^{2}\ge p_{n}+p_{m}$
and $2(p_{n}+p_{m})\ge(\sqrt{p_{m}}+\sqrt{p_{n}})^{2}$ one immediately
sees that \begin{equation}
\frac{ds^{2}}{2}\le ds_{QCB}^{2}\le ds^{2}.\label{ineq:metric}\end{equation}
 This relation shows that, as far as divergent behavior is concerned,
the Bures and the Chernoff bound metric are equivalent i.e., one metric
diverges iff the other does. On the other hand in the metric approach
to QPTs the identification of divergences of the {\em rescaled}
metric tensor and their study plays the central role \cite{DG-qpt}.
Therefore one expects the two distinguishability measures to convey
equivalent information about the location of the QPTs. Though most
of the calculations that are reported in this paper could be easily
extended to the Chernoff bound metric, here we will limit ourselves
to the analysis of the Bures metric (\ref{eq:dist}). %A thorough study of the Chernoff bound
%metric (\ref{QCB-metric} in relation to quantum criticality and at
%finite temperature will be provided elsewhere \cite{abasto}.

\section{Thermal states}

From now on we specialize our analysis to the case of thermal states.
If the Hamiltonian smoothly depends on a set of parameters, denoted
by $\lambda,$ living in same manifold ${\cal M}$ one has the smooth
map $(\lambda,\beta)\rightarrow\rho(\beta,\lambda):=Z^{-1}e^{-\beta H(\lambda)},\,(Z={\rm {tr}}e^{-\beta H}).$
What we are going to study in this paper is basically the pull-back
onto the $(\lambda,\beta)$ plane of the Bures metric through this
map. This is the obvious finite-temperature extension of the ground-state
approach of Ref. \cite{DG-qpt}.

We start by studying the Bures distance when $T\neq0$ is fixed and
for infinitesimal variations of the Hamiltonian's parameters $\lambda$.
Notice first that $\rho=Z^{-1}\sum_{n}e^{-\beta E_{n}}|n\rangle\langle n|$
where $E_{n}$ and $|n\rangle$ are the eigenvalues and eigenvectors
of the Hamiltonian operator $H.$ With a standard reasoning, by differentiating
the Hamiltonian eigenvalue equation one finds that $\langle i|dj\rangle=\langle i|dH|j\rangle/(E_{i}-E_{j})$.
Moreover one easily sees that $dp_{i}=d(e^{-\beta E_{i}}/Z)=-\beta p_{i}(dE_{i}-(\sum_{j}dE_{j}p_{j})),$
therefore the first term in equation (\ref{start}) can be written
as $\beta^{2}/4\sum_{i}p_{i}(dE_{i}^{2}-\langle dE\rangle)^{2}$ where
$\langle dE\rangle_{\beta}:=\sum_{j}dE_{j}p_{j})$. This means that
the Fisher-Rao distance is expressed by the thermal variance of the
diagonal observable $dH_{d}:=\sum_{j}dE_{j}|j\rangle\langle j|$ times
the square of the inverse temperature. Summarizing
\begin{eqnarray}
ds_{B}^{2} & = & \frac{\beta^{2}}{4}(\langle dH_{d}^{2}\rangle_{\beta}-\langle dH_{d}\rangle_{\beta}^{2})+\nonumber \\
 &  & \frac{1}{2}\sum_{n\neq m}\left|\frac{\langle
 n|dH|m\rangle}{E_{n}-E_{m}}\right|^{2}\frac{(e^{-\beta E_{n}}-e^{-\beta
 E_{m}})^{2}}{Z(e^{-\beta E_{n}}+e^{-\beta
 E_{m}})}.\label{dist-therma}
\end{eqnarray} 
The two terms correspond to the first and second term of (\ref{start})
respectively and they depend on $\beta$ and on the other parameters
of the Hamiltonian. For example, when a single parameter $h$ is considered,
the Bures distance defines a simple metric that can be expressed in
term of the classical and non-classical part \be g_{hh}(h,\beta)=
g_{hh}^c(h,\beta)+g_{hh}^{nc}(h,\beta)\label{gcommnoncomm} \ee
such that $ds_{B}^{2}=g_{hh}(h,\beta)dh^{2}.$ Let us now explore
the behavior of the Bures distance in presence of infinitesimal variations
of both the temperature ($\beta$ variations) and a field $h$ in
the Hamiltonian. It is easy to see that the variation of $\beta$
only affects the Fisher Rao classical term in (\ref{start}). In fact
the variation $dH$ in (\ref{dist-therma}), or analogously the variations
$\ket{dm}$ in (\ref{start}), are taken with respect to $h$ only.
The calculations can be summarized as follows. We first have to expand
the $dp_{n}$ as $dp_{n}=(\partial_{\beta}p_{n})d\beta+(\partial_{h}p_{n})dh$.
We have that \[
(\partial_{\beta}p_{n})d\beta=p_{n}\left[\langle H\rangle-E_{n}\right]d\beta\]
 and \[
(\partial_{h}p_{n})dh=\beta \, p_{n}\left[\langle\partial_{h}H_{d}\rangle-\partial_{h}E_{n}\right]dh\]
 where $E_{n}=E_{n}(h)$. The complete classical term of the Bures
distance can be written expanding $(dp_{n})^{2}=(\partial_{h}p_{n}dh)^{2}+(\partial_{\beta}p_{n}d\beta)^{2}+2\partial_{\beta}p_{n}\partial_{h}p_{n}d\beta dh$,
and summing over $n$. We thus have three different contributions:

\bae \frac{1}{4}\sum_{n}\frac{(dp_{n})^{2}}{p_{n}}
& =& \frac{1}{4}\left\{ \rule{0mm}{4mm} \left[ \langle H^{2}\rangle-\langle
H\rangle^{2}\right]d\beta^{2}\right.\nonumber \\
 & & +\beta^{2}\left[\langle\left[(\partial_{h}H)_{d}\right]^2
\rangle-\langle(\partial_{h}H)_{d}\rangle^{2}\right]dh^{2}\nonumber
\\
 & & \left.+2\beta\left[\langle H(\partial_{h}H)_{d}\rangle-\langle(\partial_{h}H)_{d}\rangle\langle
H\rangle\right]d\beta dh \rule{0mm}{4mm}\right\} \nonumber \\
 \label{comm_h_beta} \eae These terms correspond to the elements
of the metric $g_{\beta,\beta},\ \ g_{h,\beta},\ \ g_{h,h}^{c}$ respectively.
The full metric can be written once one calculates the non classical
term in (\ref{start}). The infinitesimal Bures distance can then
be written in terms of the $2\times2$ metric tensor $g$ as :

\be ds^{2}=\left(dh,d\beta\right) g \left(\begin{array}{c}
dh\\
 d\beta\end{array}\right),\quad{}g=\left(\begin{array}{cc}
g_{hh} & g_{h\beta}\\
 g_{h\beta} & g_{\beta\beta}\end{array}\right).\label{g_t&h}
\ee where again $g_{hh}(h,\beta)=g_{hh}^{c}(h,\beta)+g_{hh}^{nc}(h,\beta).$

It is at this point interesting to check whether, for $\beta\rightarrow\infty,$
one recovers the known results for ground-state (pure) fidelity and
metric tensor. In order to do that we will consider separately the
classical and non-classical term in (\ref{start}). In fact $\|\rho(\beta)-\rho(\infty)\|_{1}=(1-p_{0})+\sum_{n>0}p_{n}\le2\sum_{n>0}e^{-\beta(E_{n}-E_{0})};$
from which one sees that, for finite-dimensional systems, the thermal
density matrix converges (in trace norm) exponentially fast to the
projector over the ground state $|0\rangle.$ Then it follows that
all the expectations values will converge exponentially fast to their
zero-temperature limits: $|{\rm {tr}(A\rho(\beta))-{\rm {tr}(A\rho(\infty))|\le\| A\|\|\rho(\beta)-\rho(\infty)\|_{1},}}$
this in turn guarantees that the covariances of diagonal operators
in the Fisher-Rao term (\ref{comm_h_beta}) are vanishing (since e.g.,
$dH_{d}$ is diagonal $\langle dH^{2}\rangle_{\infty}=\langle dH\rangle_{\infty}^{2}.$)
in the zero-temperature limit. In the infinite dimensional case the
convergence to zero of this term will typically be only algebraic
in the region where the smallest excitation gap is small compared
to the temperature, whereas it will be exponential elsewhere. The
overall convergence behavior of the classical term for $\beta\rightarrow\infty$
depends now on the detailed interplay between the decay of covariances
we just discussed in (\ref{comm_h_beta}) and the divergence of the
powers of $\beta$ in front of them. An analysis of the zero-temperature
limit of these terms will be provided later for the quantum Ising
model. We will see that all the classical terms vanish in the zero
temperature limit but at the critical value of the parameter. As far
as the second non-classical term in (\ref{start}) is concerned one
has just to notice that from $\lim_{\beta\to\infty}p_{n}(\beta)=\delta_{n,0}$
it follows that the only contributions will come from the elements
involving the ground-state i.e., $\langle0|dj\rangle=\langle0|dH|j\rangle/(E_{j}-E_{0}).$
This completes the remark.

Before moving to the next sections, where we will specialize the previous
results to the particular case of the Quantum Ising model, we would
like to notice that the variation of the Bures distance with temperature
only, given by the element $g_{\beta\beta}$ of the metric, is precisely
proportional to the \emph{specific heat} $c_{v}$ \cite{DG-qpt},
i.e. \[
ds_{B}^{2}=\frac{d\beta^{2}}{4}(\langle H^{2}\rangle_{\beta}-\langle H\rangle_{\beta}^{2})=\frac{d\beta^{2}}{4}T^{2}c_{v}.\]
 This simple fact was already observed in \cite{DG-qpt} and \cite{gu}
and provides, we believe, a neat connection between quantum-information
theoretic concept, geometry and thermodynamics.

\section{Quasi free fermions}

In this section we specialize the study of the behavior of the Bures
metric to systems of quasi-free fermions when one has the variation
of one parameter $h$ of the Hamiltonian and of the temperature $T$.
The results that we present here are a finite-temperature generalization
of those given in Refs \cite{za-co-gio} and \cite{co-gio-za} and
directly related to the mixed-state fidelity ones reported in \cite{zhong-guo}.

The quasi-free Hamiltonians we consider are given, after performing
a suitable Bogoliubov transformation, by \begin{equation}
H=\sum_{\nu}\Lambda_{\nu}\eta_{\nu}^{\dagger}\eta_{\nu},\label{FF}\end{equation}
 where $\Lambda_{\nu}>0$ and $\eta_{\nu}$ denote the quasi-particle
energies and annihilation operator respectively. One has that $\nu$
is a suitable quasi-particle label, that for translationally invariant
systems amounts to a linear momentum; the ground state is the vacuum
of the $\eta_{\nu}$ operators i.e., $\eta_{\nu}\ket{GS}=0,\ \forall\ \nu$.
The dependence on the parameter $h$ is both through the $\Lambda_{\nu}$'s
and the $\eta_{\nu}$'s.

We now derive the explicit general form of the Bures distance (\ref{eq:dist})
starting from the classical part (\ref{comm_h_beta}). We observe
that the (many-body) Hamiltonian eigenvalues are given by $E_{j}=\sum_{\nu}n_{\nu}\Lambda_{\nu}$
where the $n_{\nu}$'s are fermion occupation numbers i.e., $n_{\nu}=0,1.$
Therefore we have that $dE_{j}=\sum_{\nu}n_{\nu}d\Lambda_{\nu}$ and
$\langle dE_{j}\rangle_{\beta}=\sum_{\nu}\langle n_{\nu}\rangle_{\beta}d\Lambda_{\nu}$
where the averages are easy to compute since the probability distribution
of the $dE_{j}$ factorizes over the $\nu$'s. Furthermore, $\langle n_{\mu}n_{\nu}\rangle_{\beta}-\langle n_{\mu}\rangle_{\beta}\langle n_{\nu}\rangle_{\beta}=\delta_{\mu\nu}\langle n_{\nu}\rangle_{\beta}\left(1-\langle n_{\nu}\rangle_{\beta}\right)$
and we can thus write: \bae \frac{1}{4}\sum_{n}\frac{(dp_{n})^{2}}{p_{n}}
& = & \frac{1}{4}\sum_{k}\langle n_{k}\rangle\left(1-\langle
n_{k}\rangle\right)\times \nonumber \\
 & & \left\{ \Lambda_{k}^{2}\ \ d\beta^{2}\right.\nonumber
\\
 & & +\beta^{2}(\partial_{h}\Lambda_{k})^{2}\ \ dh^{2}\nonumber
\\
 & & \left.+2\beta\Lambda_{k}\partial_{h}\Lambda_{k}\ \ d\beta\ dh\right\}.
\eae The term in $dh^{2}$ is the classical term due to the infinitesimal
variations of the parameters of the Hamiltonian at fixed $T$ and
it corresponds to the variance, see (\ref{dist-therma}), $var(H_{d})=\sum_{\nu}\langle n_{\nu}\rangle_{\beta}(1-\langle n_{\nu}\rangle_{\beta})d\Lambda_{\nu}^{2}.$
Since we are dealing with independent free-fermions one has that $\langle n_{\nu}\rangle_{\beta}=(\exp(\beta\Lambda_{\nu})+1)^{-1},$
whence \begin{equation}
ds_{c}^{2}=\frac{\beta^{2}}{16}\sum_{\nu}\frac{(\partial_{h}\Lambda_{k})^{2}}{\cosh^{2}(\beta\Lambda_{\nu}/2)}dh^{2}\label{FF-diag}\end{equation}
 In order to compute the non-classical part of Eq. (\ref{start}),
one has to explicitly consider the eigenvectors of (\ref{FF}). Following
the notation of Ref. \cite{za-co-gio} one has $|m=\{\alpha_{\nu},\alpha_{-\nu}\}_{\nu>0}\rangle=\otimes_{\nu>0}|\alpha_{\nu},\alpha_{-\nu}\rangle$
where \begin{eqnarray*}
|0_{\nu}0_{\nu}\rangle & = & \cos(\theta_{\nu}/2)|00\rangle_{\nu,-\nu}-\sin(\theta_{\nu}/2)|11\rangle_{\nu,-\nu},\\
|0_{\nu}1_{-\nu}\rangle & = & |01\rangle_{\nu,-\nu},\quad|1_{\nu}0_{-\nu}\rangle=|10\rangle_{\nu,-\nu},\\
|1_{\nu}1_{\nu}\rangle & = & \cos(\theta_{\nu}/2)|11\rangle_{\nu,-\nu}+\sin(\theta_{\nu}/2)|00\rangle_{\nu,-\nu}.\end{eqnarray*}
 We assume now that parameter dependence is only in the angles $\theta_{\nu}$'s
(this assumption holds true for all the translationally invariant
systems). It is easy to see from the above factorized form that the
only non vanishing matrix elements $\langle n|dm\rangle$ are given
by $\langle0_{\nu}0_{-\nu}|d|1_{\nu}1_{-\nu}\rangle=d\theta_{\nu}/2$
and that the thermal factor $(p_{n}-p_{m})^{2}/(p_{n}+p_{m})$ has
the form $\sinh^{2}(\beta\Lambda_{\nu})/[(\cosh(\beta\Lambda_{\nu})+1)(\cosh(\beta\Lambda_{\nu})]=(\cosh(\beta\Lambda_{\nu})-1)/\cosh(\beta\Lambda_{\nu}).$
Putting all together one finds \begin{equation}
ds_{nc}^{2}=\frac{1}{4}\sum_{\nu>0}\frac{\cosh(\beta\Lambda_{\nu})-1}{\cosh(\beta\Lambda_{\nu})}(\partial_{h}\theta_{\nu})^{2}dh^{2}\label{FF-off}\end{equation}
 We finally note that the two elements (\ref{FF-diag}) and (\ref{FF-off})
define the metric element (\ref{gcommnoncomm}).The results of this
section can be applied to any quasi-free fermionic model (\ref{FF}).

\section{Quantum Ising Model}

We are now going to discuss in some detail the behavior of the metric
tensor for a paradigmatic example in the class of quasi-free fermionic
models, the Ising model in transverse field. The model is defined
by the Hamiltonian\begin{equation}
H=-\sum_{j}\sigma_{j}^{x}\sigma_{j+1}^{x}+h\sigma_{j}^{z}.\label{eq:ising_model}\end{equation}
 At $T=0$ this system undergoes a quantum phase transition for $h=1$
. For $h<1$ the system is in an ordered phase as the correlator $\langle\sigma_{1}^{x}\sigma_{r}^{x}\rangle_{T=0}$
tends to a non zero value: $\lim_{r\rightarrow\infty}\langle\sigma_{1}^{x}\sigma_{r}^{x}\rangle_{T=0}=\left(1-h^{2}\right)^{1/4}$.
The excitations in this region are domain walls in the $\sigma^{x}$
direction. Instead for $h>1$ the magnetic field dominates, and excitations
are given by spin flip over a paramagnetic ground state. The transition
point $h=1$ is described by a $c=1/2$ conformal field theory,which
implies that means that the dynamical exponent $z=1$; the correlation
function exponent is $\nu=1$. As is well known \cite{sachdev}, a
signature of the ground state phase diagram remains at positive temperature.
In the {\em quasi classical} region $T\ll\Delta$, where $\Delta=\left|1-h\right|$
is the lowest excitation gap, the system can be described by a diluted
gas of thermally excited quasi-particles, even if the nature of the
quasi-particles is different at the different sides of the transition.
Instead in the {\em quantum critical} region $T\gg\Delta$ the
mean inter-particle distance becomes of the order of the quasi-particle
de Broglie wavelength and thus quantum critical effects dominate and
no semiclassical theory is available. In each of the above described
regions of the $\left(h,T\right)$ plane the system displays very
different dynamical as well thermodynamical properties. For example,
in the quantum critical region the specific heat approaches zero linearly
with the temperature (this is in fact a general feature of all conformal
field theories), whereas in the quasi-classical regions the approach
is exponentially fast.

\subsection{Bures metric tensor in the $\left(h,T\right)$ plane}

We now investigate whether the signature of the physically different
regions can be revealed by analyzing the elements of the metric tensor
defined by the Bures distance. We begin by studying the temperature
dependence of the metric tensor when only the external field is varied
i.e., the term $g_{hh}(h,T)$, see Eq.~(\ref{gcommnoncomm}). The
Hamiltonian (\ref{eq:ising_model}) is equivalent to a quasi-free
fermionic model, and following our previous notation one has $\epsilon_{k}=\cos\left(k\right)-h$,
$\Delta_{k}=\sin\left(k\right)$, $\Lambda_{k}=\sqrt{\epsilon_{k}^{2}+\Delta_{k}^{2}}$
and $\tan\left(\vartheta_{k}\right)=\Delta_{k}/\epsilon_{k}$. Using
formulae (\ref{FF-diag}) and (\ref{FF-off}) it is straightforward
to write (\ref{gcommnoncomm}). After rescaling $g\rightarrow g/L$
and passing to the thermodynamic limit we obtain \begin{eqnarray*}
g_{hh}^{c} & = & \frac{\beta^{2}}{16\pi}\int_{-\pi}^{\pi}\frac{1}{\cosh\left(\beta\Lambda_{k}\right)+1}\frac{\epsilon_{k}^{2}}{\Lambda_{k}^{2}}dk\\
g_{hh}^{nc} & = & \frac{1}{8\pi}\int_{-\pi}^{\pi}\frac{\cosh\left(\beta\Lambda_{k}\right)-1}{\cosh\left(\beta\Lambda_{k}\right)}\frac{\Delta_{k}^{2}}{\Lambda_{k}^{4}}dk\end{eqnarray*}

The integrals are better evaluated by transforming momentum integration
to energy integration in a standard way. As previously noticed, on
general grounds, the classical term $g_{hh}^{c}$ vanishes when the
temperature goes to zero. In the quantum-critical region $\beta\Delta\approx0$,
and one obtains the following low temperature expansion:\[
g_{hh}^{c}=\frac{\pi}{96h^{2}}T+O\left(T^{2}\right).\]
 Instead in the quasi-classical region where $\beta\Delta\gg1$ the
fall-off to zero is exponential. With a saddle point approximation
one obtains\[
g_{hh}^{c}=\sqrt{\frac{\Delta}{32\pi h}}T^{-3/2}e^{-\Delta/T}+\mathrm{lower}\,\,\mathrm{order}.\]

We now analyze the scaling behavior of the non-classical term of
the metric $g$. From the results of \cite{za-pa,za-co-gio} it is
known that the geometric tensor at zero temperature diverges as $\Delta^{-1}$
when $\Delta\rightarrow0$. The non classical term matches this ground
state behavior from positive temperature. Indeed, in the quantum-critical
region the integral is well approximated by \[
g_{hh}^{nc}\approx\frac{1}{8\pi h{}^{2}}\int_{0}^{2\beta}\frac{\cosh\left(x\right)-1}{\cosh\left(x\right)}\frac{\sqrt{\left({4\beta}^{2}-x^{2}\right)}}{x^{2}}dx.\]
 For large $\beta$ (low temperatures) this expression can be Laurent
expanded and the resulting integrals can be re-summed using residuum
theorem, giving\begin{equation}
g_{hh}^{nc}=\frac{1}{h^{2}}\left[\frac{\mathcal{C}}{\pi^{2}}T^{-1}-\frac{1}{16}+O\left(T\right)\right],\label{eq:goff_crit}\end{equation}
 where $\mathcal{C}$ is Catalan's constant $\mathcal{C}=0.915966\ldots$.

We would like to point out that the behavior of the metric tensor
in the quasi-critical region, can be inferred from dimensional scaling
analysis in much the same spirit as was done in \cite{lor} for the
zero temperature metric tensor. From Eq.~(\ref{dist-therma}) we
see that the scaling dimension of $g_{hh}^{nc}$ is $\Delta_{nc}=2\Delta_{V}-2z-d$,
where $\Delta_{V}$ is the scaling dimension of the operator $dH$,
$z$ is the dynamical exponent and $d$ is the spatial dimensionality.
Following \cite{lor} ($\beta$ now plays the role of the length)
we obtain \begin{equation}
g_{hh}^{nc}\sim T^{\Delta_{nc}/z}\label{eq:g_hh}\end{equation}
 In the present case, $z=d=\Delta_{V}=1$ (the scaling dimension of
$\sigma_{i}^{z}$ --a free fermionic field-- is one) which agrees
with (\ref{eq:goff_crit}).

We now pass to analyze the behavior of $g_{hh}^{nc}$ in the quasi-classical
region i.e. when $\beta\Delta\gg1$. In this case the {}``temperature''
part of the integral is never effective, i.e. one has $\frac{\cosh\left(\beta\Lambda\right)-1}{\cosh\left(\beta\Lambda\right)}\approx1$,
so it is quite clear that, in first approximation, one recovers the
zero temperature result first given in \cite{za-pa} which we re-write
here as an energy integral \begin{equation}
g_{hh}^{nc}=\frac{1}{8\pi h^{2}}\int_{\left|1-h\right|}^{\left|1+h\right|}\frac{\sqrt{\left[\left(h-1\right)^{2}-\omega^{2}\right]\left[\omega^{2}-\left(1+h\right)^{2}\right]}}{\omega^{3}}d\omega,\label{eq:goffzero}\end{equation}
 and we assumed $h>0$. For small values of the gap, ---hence we are
in a situation where we consider first the limit $T\rightarrow0$
and then $\Delta\rightarrow0$--- we observe the following divergence\[
g_{hh}^{nc}(T=0,\Delta\rightarrow0)\approx\frac{1}{16\Delta},\]
 which is a result also reported in \cite{za-pa,za-co-gio}. Instead
when the gap is large, ---so that we are necessary on the $h>1$ side---
we can approximate the radical in Eq.~(\ref{eq:goffzero}) with an
ellipse centered at $\left(h,0\right)$ with semi-axes $r_{x}=1$
and $r_{y}=2h$, that amounts to write $\sqrt{\left[\left(h-1\right)^{2}-\omega^{2}\right]\left[\omega^{2}-\left(1+h\right)^{2}\right]}\approx2h\sqrt{1-\left(\omega-h\right)^{2}}$.
In this case the integral gives\[
g_{hh}^{nc}\left(T=0,\Delta\gg1\right)\approx\frac{1}{8h^{5/2}\left(h-1\right)^{3/2}}\approx\frac{1}{8\Delta^{4}}.\]
 Again, by doing a saddle point approximation one realizes that the
zero temperature results are approached exponentially fast with the
temperature, more precisely one has\[
g_{hh}^{nc}\left(\beta\Delta\gg1\right)=g_{hh}^{nc}\left(T=0\right)-\mathrm{const.}\times T^{3/2}e^{-\Delta/T}.\]

We now extend our analysis to the other terms of metric tensor (\ref{g_t&h}).
When we consider the case in which both the temperature and the field
$h$ are varied two new matrix elements come into play: \begin{eqnarray*}
g_{TT} & = & \frac{\beta^{4}}{16\pi}\int_{-\pi}^{\pi}\frac{\Lambda_{k}^{2}}{\cosh\left(\beta\Lambda_{k}\right)+1}dk\\
g_{hT} & = & \frac{\beta^{3}}{16\pi}\int_{-\pi}^{\pi}\frac{\epsilon_{k}}{\cosh\left(\beta\Lambda_{k}\right)+1}dk.\end{eqnarray*}

Let us first comment on the behavior observed at very low temperature.
In the quasi-classical region ($\Delta\gg T$) all matrix elements
of $g$ tend to zero except for $g_{hh}^{nc}$. This is a general
feature and is due to the fact that these terms are absent in the
zero temperature expression. As previously stated the fall-off to
zero is exponential, and in particular, for the model in exam, we
have that \bae g_{Th} & \approx & T^{-5/2}e^{-\Delta/T}\,,\quad{}T\ll\Delta
\nonumber\\
 g_{TT} & \approx & T^{-7/2}e^{-\Delta/T}. \eae Let us now
look at the quantum critical region $T\gg\Delta$, small temperature.
The mixed term tends to a constant: \be g_{hT}=\frac{\pi}{48}+O\left(T^{2}\right).
\ee Instead $g_{TT}$ must diverge at zero temperature, as it has
to match with the diverging behavior observed in the ground state
\cite{za-pa}. For the diagonal term $g_{TT}$ one has \be g_{TT}=\frac{T^{-2}}{4}c_{v}=\frac{\pi}{24}\frac{1}{T}+O\left(T\right),\quad{}T\gtrsim\Delta.
\label{eq:gtt_quasi_critical} \ee We note in passing that this result
agrees with the one for the specific heat obtained for general conformal
theories \cite{affleck86} $c_{v}=\left(\pi cT\right)/\left(3v\right),\quad\mbox{as }T\rightarrow0$,
since in our case the velocity $v$ is one and the conformal charge
$c$ is one-half. We thus see that, in the present case, both $g_{hh}$
and $g_{TT}$ diverge as $T^{-1}$. This has not to be the case in
general, indeed at any quantum critical point described by a conformal
field theory, $g_{TT}$ will diverge as $T^{-1}$ whereas the behavior
of $g_{hh}^{nc}$ is dictated by Eq.~(\ref{eq:g_hh}).

In this section we have analyzed the behavior of all the elements
of the geometric tensor $g$. The result of this analysis allows one
to conclude that indeed, at least for the specific model studied,
{\em the quantum critical and quasi classical regions can be clearly
identified in terms of the markedly different temperature behavior
of the geometric tensor $g$}.

\subsection{Directions of maximal distinguishability}

The analysis carried out in the previous section can be further deepened
by studying some useful quantities that can be derived by the analysis
of the metric tensor $g$. Indeed, we will see that these quantities
allow to give a finer description of the behavior of the system in
the plane $(h,T)$ and to reveal new unexpected features. We first
start by noticing that at each point $\left(h,T\right)$ the eigenvectors
of the metric tensor $g$ define the directions of maximal and minimal
growth of the line element $ds_{B}^{2}$. Hence the vector field $\vec{v}_{M}(h,T)$
given by the eigenvector of $g$ related to the highest eigenvalue
$\lambda_{M}$, defines at each point of the $(h,T)$ plane the direction
along which the fidelity decreases most rapidly: the latter represents
the direction of highest distinguishability between two nearby Gibb's
states.

We now focus our analysis on the study of the vector field $\vec{v}_{M}(h,T)$
in the specific case of the quantum Ising model, see Fig.~\ref{fig:vector-field}.
We first observe that there clearly are some interesting features
for small temperatures, that reflect the analysis previously carried
out on the metric elements of $g$. On one hand, in the quasi-classical
region, when $h\simeq0$ we have that the direction of highest fidelity
drop, is parallel to the $h$ axis. This reflects the fact that, in
this region, all the elements of $g$ tend to zero except for the
term $g_{hh}^{nc}$. On the other hand, in the quasi critical region,
the direction of highest fidelity drop is parallel to the $T$ axis.
Again, this feature can be linked to the previous analysis of the
terms (\ref{eq:goff_crit}) and (\ref{eq:gtt_quasi_critical}). Indeed
both $g_{hh}$ and $g_{TT}$ diverge as $T^{-1}$, but, as $\left(\mathcal{C}/\pi^{2}\right)/\left(\pi/24\right)=0.70\ldots<1$,
$g_{TT}$ eventually becomes bigger and thus the direction of highest
distinguishability turns parallel to the $T$ axis.

\begin{center}%
\begin{figure}
\begin{centering}\includegraphics[width=6cm,keepaspectratio]{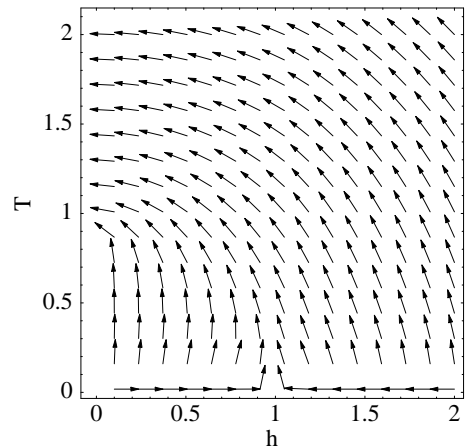}\par\end{centering}

\caption{Vector field of the eigenvector associated to the highest eigenvalue
of $g$, in the plane $\left(h,T\right)$ for the Ising model in transverse
field.\label{fig:vector-field}}
\end{figure}
\par\end{center}

We proceed in our description of the phase diagram through the introduced
vector field by examining what happens at the $h=0$ axis. Here one
has the impression that a kind of singular point appear around $T\approx1$.
The reason for this is that at $h=0$ the system becomes the purely
classical Ising model, which possess only classical behavior at any
temperature. This implies that the quantum-critical region cannot
extend over this line. As the dispersion $\Lambda_{k}$ is flat, it
is straightforward to write down the metric tensor on the $h=0$ line.
It turns out that $g$ is completely diagonal meaning that eigenvectors
are parallel to the $\left(h,T\right)$ axes. One sees that for
$0.852<T<\infty$, 
$g_{hh}>g_{TT}$ then for $0.101<T<0.852$, $g_{TT}>g_{hh}$ and then
finally, at very small temperature, quantum fluctuations dominates
and for $0<T<0.101$, $g_{hh}>g_{TT}$. The appearance of the purely
classical Ising line at $h=0$, which forbids the quantum critical
phase extend over this line, is related to the fact that the model
(\ref{eq:ising_model}) is invariant under the $\mathbb{Z}_{2}$ symmetry
$h\rightarrow-h$. This in turns implies that the phase diagram is
mirror symmetric around the $h=0$ line and that there is another
quantum critical point at $h=-1$. The physical consequence is that
the semiclassical ordered region is much smaller than one would think
and the actual phase diagram is much like in Figure \ref{fig:phase-diagram}. 

\begin{center}%
\begin{figure}[h]
\begin{centering}\includegraphics[width=7cm,keepaspectratio]{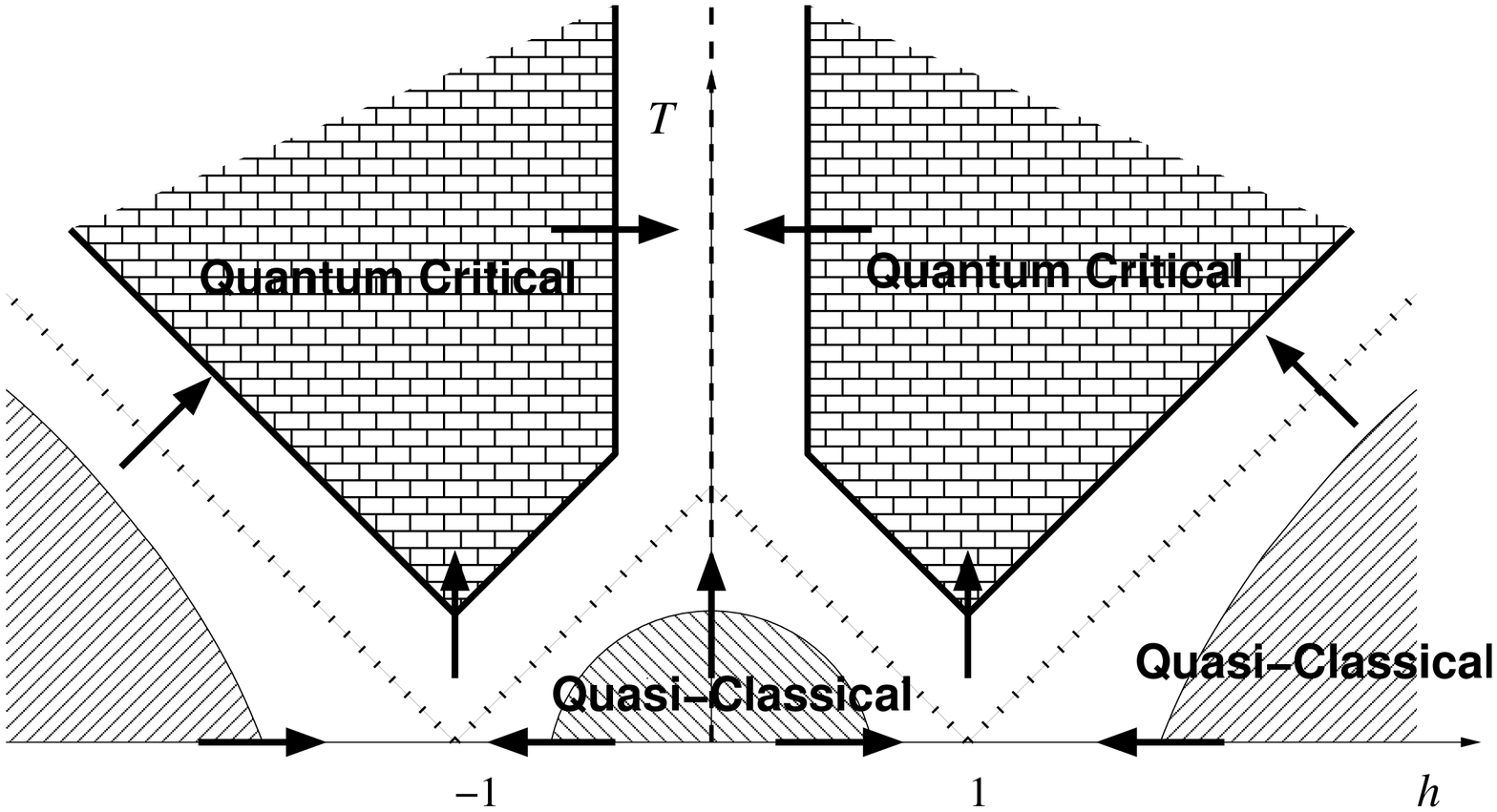}\par\end{centering}

\caption{Phase diagram of the Ising model in transverse field taking into
account both critical points at $h=\pm1$, and the purely classical
Ising line $h=0$. The arrow indicates the direction of highest fidelity
decrease (the direction of the arrows is conventional, but fixed once
for all)\label{fig:phase-diagram}}
\end{figure}
\par\end{center}

Finally we now discuss another feature that can observed by studying
$\vec{v}_{M}(h,T)$. As one can see in Fig.~\ref{fig:vector-field},
along the line $T=h\gtrsim1$ the vector field becomes parallel to
the vector $\vec{w}=(-1,1)$. It turns out that this feature can be
understood analytically by studying the behavior of the metric tensor
$g$ when $|h|\gg1$. Indeed, by evaluating the dominant part of the
various metric elements on the line $T=h=t\gg1$, one sees that all
the Fisher-Rao terms decay as $t^{-2}$ while $g_{hh}^{nc}\sim t^{-4}$,
and, what is most surprising, all matrix elements tend to have the
same value in magnitude. This feature can be understood by simply
observing that when $|h|\gg1$ it is only the classical term proportional
to the external magnetic field of the Quantum Ising Hamiltonian that
survives i.e., $H\simeq h\sum_{i}\sigma_{i}^{z}$. The density matrix
of the system can be written as $\rho(h,T)=\exp{(-h\sum_{i}\sigma_{i}^{z}/T)}/Z$;
in this approximation the only non zero terms of the metric are the
Fisher-Rao ones and all the covariances that define these terms, see
(\ref{comm_h_beta}), coincide with $var(H)$. Thus, in
the limit $|h|\gg1$ the Bures distance reads $ds_{B}^{2}=var(H)[dT^{2}/T^{2}-hdTdh/T^{3}+h^{2}dh^{2}/T^{4}]$.
If now one chooses the particular case $T=h=t$ and evaluates the
density $g/L$ one finds that \[
g\left(t,t\right)=\frac{t^{-2}}{16\cosh^{2}\left(1/2\right)}\left(\begin{array}{cc}
1 & -1\\
-1 & 1\end{array}\right)+O\left(t^{-3}\right).\]

Thus, one has that on the line $T=h\gg1$ the only non zero eigenvalue
is $2 \, var(H) /(Lt^{2})$ and it corresponds to the eigenvector
$\vec{w}=(-1,1)$. In this approximation, that amounts to neglecting
the term $g_{hh}^{c}\sim t^{-4}$, when moving along the line $T=h\gg1$
i.e., along the direction defined by $\vec{w}^{\perp}$, no changes
in the state of the system occur.

\begin{center}%
\begin{figure}
\begin{centering}\includegraphics[width=6cm,keepaspectratio]{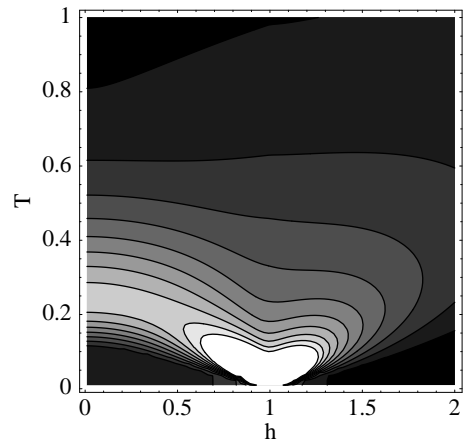} \par\end{centering}

\caption{Contour plot of the highest eigenvalue of $g$, in the plane $\left(h,T\right)$
for the Ising model in transverse field.\label{fig:maxeigen1}}
\end{figure}
\par\end{center}
\subsection{Crossover and metric tensor $g$}

We finally present some preliminary results related to the intriguing
possibility of determining the crossover lines between the quasi-classical
and quasi-critical region (\ref{eq:ising_model}) through the analysis
of the elements of metric tensor $g$ and the induced Gaussian curvature
\cite{Nak} in the plane $\left(h,T\right)$.
The capability
of the highest (in modulus) eigenvalue of the $g$ and of the Gaussian
curvature induced by the metric to capture, in terms of divergencies
or discontinuities, the existence of QPTs has been already tested in
\cite{co-gio-za} and \cite{DG-qpt}.  Here we would like to test whether
these quantities are able to identify the crossover between the quasi-classical
and quantum-critical region. Notice that the curvature of the Bures
metric in the case of squeezed states has been studied in \cite{parao} and an
operational interpretation attempted. It is also worthwhile to stress
that the so-called thermodynamical curvature plays a central role
in the geometrical theory of classical phase transition developed 
by Ruppeiner and coworkers \cite{ruppy}. 

As already pointed out, at each point
$\left(h,T\right)$ the vector field $\vec{v}_{M}(h,T)$ defines the
direction of highest distinguishability between two nearby Gibb's
states. The degree of distinguishability along this direction is quantified
by the maximal eigenvalue $\lambda_{M}\left(h,T\right)$. Since the
quasi-classical and quantum-critical regions are characterized by
significantly different physical properties, it is natural to investigate
whether the change of the latter, in spite of not involving a phase
transition, could be revealed by our measures of statistical distinguishability
and by the related functionals.

We now give a descriptive analysis of the raw data. In figure \ref{fig:maxeigen1},
we have plotted the contour plot of the maximal eigenvalue of $g$.
The main feature is the presence for $T>0$ of two patterns of high
distinguishability (white) that separate the regions ($h<1,T\lesssim0.25)$
and ($h>1,T\lesssim0.25)$ from the rest of the diagram. Thus, the
first information that can be drawn is that a change of parameters
inside these regions implies a small change in the statistical properties
of the corresponding ground states. On the contrary, if one varies
$h$ and $T$ and moves from these regions towards the center of the
diagram, for example moving along the integral lines of $\vec{v}_{M}(h,T)$
, the statistical properties of the state necessarily have to significantly
change. One can see that, the \char`\"{}transition\char`\"{} lines
between the different regions can be extrapolated numerically by tracing
the \char`\"{}ridge\char`\"{} lines of the two patterns of high distinguishability.
It turns out that the \textit{same} result can be achieved by looking
at the lines where the Gaussian curvature of $g$ changes sign, see
figure \ref{fig:curvature}. For example, when $h>1$, one can see
that along the determined transition line, $T$ has a linear dependence
on $h-1$.

This preliminary descriptive analysis seems thus to indicate that
a neat distinction between the quasi-classical regions (characterized
by a negative curvature) and the quantum-critical (characterized by
a positive curvature) can be made on the basis of study of the metric
$g$. This is indeed the first time that the use of the fidelity,
and of the related functionals, allows to identify the crossover between
two distinct phases.

\begin{center}%
\begin{figure}
\begin{centering}\includegraphics[width=9cm,keepaspectratio]{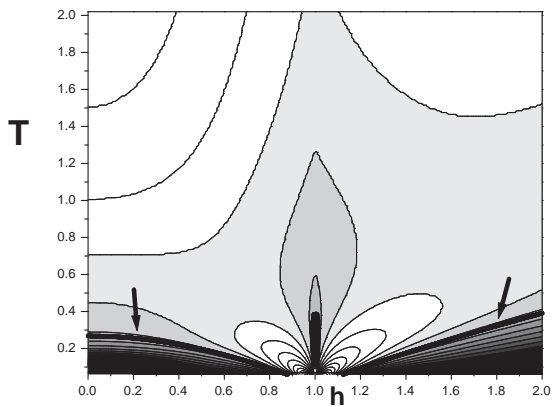} \par\end{centering}

\caption{Contour plot of the Gaussian curvature of $g$, in the plane $\left(h,T\right)$
for the Ising model in transverse field. The arrows indicate the zero
curvature lines.\label{fig:curvature}}
\end{figure}
\par\end{center}

%To summarize, the analysis of the metric tensor over the relevant
%parameter manifold $\left(h,T\right)$ allows to provide complementary
%information on the phase diagram, and can turn especially useful in
%characterizing ubiquitous cross-over lines. In the example of the
%quantum Ising model the direction of highest fidelity decrease is
%in fact perpendicular to the crossover lines (Fig.~\ref{fig:phase-diagram}).

\section{Conclusions}

In this paper we have analyzed the relation between quantum criticality,
finite temperature and the differential-geometry of the manifold of
mixed quantum states. We studied the Bures metric over the set of
thermal quantum states associated with Hamiltonians featuring a zero-temperature
quantum phase transition i.e., quasi-free fermionic systems. In particular
we focused on the study of the quantum Ising model for which we provided
a fully analytical characterization of the Bures metric tensor $g$.
Quantum critical and semiclassical regions in the temperature, magnetic
field plane can be easily identified in terms of different scaling
behavior of the components of $g$ as a function of the temperature.
Cross-over lines between the different regions can be found just by
looking at the shape of the graph of the largest eigenvalue of the
metric as a function of temperature and magnetic field. %i.e., graph of the map $(h, T)\rightarrow
%\|g\|$.
Remarkably these cross-over lines seem to be associated also with
the change of sign of the Gaussian curvature of the metric $g$.

The results presented in this paper provide further support to the
validity of the statistical-metric approach to phase transitions \cite{DG-qpt}
and clearly show that the scope of this geometrical method can be
extended to finite temperatures. The physical significance of the
curvature of the metric as well as the study of the thermal states
geometry associated with other distinguishability distances e.g.,
the quantum Chernoff bound metric, are topics deserving further investigations.
%This graph shows indeed two regions of low distingushability
%separated by a ridge over which one has a significant change of the statistical properties
%of the thermal state

\begin{acknowledgments}
The authors would like to thank for useful discussions
R. Ionicioiu and M. Paris.
\end{acknowledgments}

%%%%%%%%%%%%%%%%%%%%%%%%%%%%%%%%%%%%%%%%%%%%%%%%%%%%%%%%%%%%%%%%%%%%%%%%%%%%%

\end{document}